\documentclass[fleqn,9pt]{olplainarticle}
\usepackage{xcolor}
\usepackage{makecell}
\usepackage{pdflscape}
\usepackage{url}
\usepackage{subcaption} 
\usepackage[numbers]{natbib}
\title{Artificial Intelligence in Ovarian Cancer Histopathology: A Systematic Review}
\author[1]{Jack Breen\footnote[2]{Corresponding author - scjjb@leeds.ac.uk.}}
\author[2]{Katie Allen}
\author[3]{Kieran Zucker}
\author[2,4]{Pratik Adusumilli}
\author[2,4]{Andy Scarsbrook}
\author[3]{Geoff Hall}
\author[2]{Nicolas M. Orsi*}
\author[1]{Nishant Ravikumar*}
\affil[1]{Centre for Computational Imaging and Simulation Technologies in Biomedicine (CISTIB), School of Computing, University of Leeds, UK}
\affil[2]{Leeds Institute of Medical Research at St James's, School of Medicine, University of Leeds, UK}
\affil[3]{Leeds Cancer Centre, St James's University Hospital, Leeds, UK}
\affil[4]{Department of Radiology, St. James University Hospital, Leeds, UK}
\affil[*]{Indicates joint final authors}
\begin{abstract}
\sffamily
\subsection*{Purpose}
To characterise and assess the quality of published research evaluating artificial intelligence (AI) methods for ovarian cancer diagnosis or prognosis using histopathology data. 

\subsection*{Methods}
A search of PubMed, Scopus, Web of Science, Cochrane Central Register of Controlled Trials, and WHO International Clinical Trials Registry Platform was conducted up to \textcolor{black}{19/05/2023}. The inclusion criteria required that research evaluated AI on histopathology images for diagnostic or prognostic inferences in ovarian cancer, including primary tumours of the ovaries, fallopian tubes, and peritoneum. Reviews and non-English language articles were excluded. The risk of bias was assessed for every model that met the inclusion criteria using the Prediction model Risk Of Bias ASsessment Tool (PROBAST). Information about each model of interest was tabulated and summary statistics were reported. Based on the results, we provided recommendations to improve study design and reporting to reduce the risk of bias and improve the reproducibility of future research in the field.  The study protocol was registered on PROSPERO (CRD42022334730). PRISMA 2020 reporting guidelines were followed.

\subsection*{Results}
A total of \textcolor{black}{1573} records were identified, of which \textcolor{black}{45} were eligible for inclusion. These studies reported \textcolor{black}{80 total} models of interest, \textcolor{black}{including 37 diagnostic models, 22 prognostic models, and 21 models with other diagnostically relevant outcomes, such as tumour segmentation and cell-type classification.} 
Models were developed using 1-1375 slides from 1-\textcolor{black}{776} ovarian cancer patients. \textcolor{black}{There was a wide array of model} outcomes, \textcolor{black}{the most common being treatment response (11/80), malignancy status (10/80), stain quantity (9/80), histological subtype (7/80), and overall survival (6/80)}. Older studies used traditional machine learning (ML) models with hand-crafted features, while newer studies typically employed deep learning to automatically learn features and predict \textcolor{black}{model} outcome(s). All models were found to be at high or unclear risk of bias overall, with most research having a high risk of bias in the analysis and a lack of clarity regarding participants and predictors in the study. \textcolor{black}{Research frequently suffered from insufficient reporting and limited validation using small sample sizes}, with external validations being particularly rare.
\subsection*{Conclusion}
Limited research has been conducted on the application of AI to histopathology images for diagnostic or prognostic purposes in ovarian cancer, and none of the associated models have been demonstrated to be ready for real-world implementation. Recommendations are provided addressing underlying biases and flaws in study design, which should help inform higher-quality reproducible future research. 
Key aspects to help ensure clinical translation include more transparent and comprehensive reporting of data provenance and modelling approaches, as well as improved quantitative performance evaluation using cross-validation and external validations. 

\subsection*{Funding} Engineering and Physical Sciences Research Council and The Tony Bramall Charitable Trust
\end{abstract}
\begin{document}
\flushbottom
\maketitle
\thispagestyle{empty}

\section*{Introduction}
\sffamily

Ovarian cancer is the eighth most common malignancy in women worldwide \cite{Sung2021}. It is notoriously difficult to detect and diagnose, with ineffective screening \cite{Menon2021} and non-specific symptoms similar to those caused by menopause \cite{Ebell2016}. Encompassing primary malignant tumours of the ovaries, fallopian tubes, and peritoneum, the disease has often started to spread within the abdomen at the time of diagnosis (FIGO \cite{Berek2021} Stage 3). This typical late stage at diagnosis makes ovarian cancer a particularly deadly disease, with the 314,000 new cases diagnosed each year translating to 207,000 deaths per year globally \cite{Sung2021}. 

Most ovarian cancers are carcinomas (cancers of epithelial origin) which predominantly fall into five histological subtypes: high-grade serous, low-grade serous, clear cell, endometrioid, and mucinous. Non-epithelial ovarian cancers are \textcolor{black}{much less common} and include germ cell, sex cord-stromal, and mesenchymal tumours. Ovarian cancer subtypes differ morphologically and prognostically and have varying treatment options \cite{Kobel2008}. High-grade serous carcinoma is the most common form of ovarian cancer, accounting for approximately 70\% of all cases \cite{Prat2014}. 

Histopathology, the examination of tissue specimens at the cellular level, is the gold standard for ovarian cancer diagnosis. Pathologists typically interpret tissue stained with haematoxylin and eosin (H\&E), 
though interpretation can be a subjective, time-consuming process, with some tasks having a high level of inter-observer variation \cite{Matsuno2013, Kobel2014,  Barnard2018}. In the assessment of difficult cases, general pathologists may seek assistance from subspecialty gynaecological pathology experts, and/or use ancillary tests, such as immunohistochemistry (IHC). 
Referrals and ancillary testing can be essential to the accuracy of the diagnostic process but come at the cost of making it longer and more expensive. Worldwide, pathologists are in much greater demand than supply, with significant disparities in the number of pathologists between countries \cite{Wilson2018}, and with better-supplied countries still unable to meet demand \cite{RCPath2018}.

Traditionally, pathologists have analysed glass slides using a light microscope. However, the implementation of a digital workflow, where pathologists review scanned whole slide images (WSIs) using a computer, is becoming more common. While digital pathology uptake has likely been driven by efficiency benefits \cite{Baidoshvili2018}, it has created an opportunity for the development of automated tools to assist pathologists. These tools often aim to improve the accuracy, efficiency, objectivity, and consistency of diagnosis. Such tools could help to alleviate the global workforce shortage of pathologists, increasing diagnostic throughput and reducing the demand for referrals and ancillary tests. This is an increasingly active area of research \cite{Stenzinger2022} and, for some malignancies, these systems are starting to achieve clinical utility \cite{Raciti2022}.


In this study, we systematically reviewed all literature in which artificial intelligence (AI) techniques (comprising both traditional machine learning (ML) and deep learning methods) were applied to digital pathology images for the diagnosis or prognosis of ovarian cancer. This included research which focused on a single diagnostic factor such as histological subtype, and studies that performed computer-aided diagnostic tasks such as tumour segmentation. The review characterises the state of the field, describing which diagnostic and prognostic tasks have been addressed, and assessing factors relevant to the clinical utility of these methods, such as the risks of bias. Despite ovarian cancer being a particularly difficult disease to detect and diagnose, and the shortage of available pathologists, AI models have not yet been implemented in clinical practice for this disease. 
This review aims to provide insights and recommendations based on published literature to improve the clinical utility of future research, including reducing risks of bias, improving reproducibility, and increasing generalisability. 

\section*{Methods}
\subsection*{Literature Search}
Searches were conducted in three research databases, PubMed, Scopus and Web of Science, and two trial registries, Cochrane Central Register of Controlled Trials (CENTRAL) and the World Health Organisation International Clinical Trial Registry Platform (WHO-ICTRP). \textcolor{black}{The research databases only include journals and conference proceedings which have undergone peer review, ensuring the integrity of included research.} The initial searches were performed on 25/04/2022 and were \textcolor{black}{most recently} repeated on \textcolor{black}{19/05/2023}. The search strategy was composed of three distinct aspects - artificial intelligence, ovarian cancer, and histopathology. For each aspect, multiple relevant terms were combined using the \emph{OR} operator (e.g. ``artificial intelligence" OR ``machine learning"), and then these were combined using the \emph{AND} operator to ensure that retrieved research met all three aspects. The widest possible set of search fields was used for each search engine except for Scopus, where restrictions were imposed to avoid searching within the citation list of each article, which is not an available field in the other search engines. The terms 'ML' and 'AI' were restricted to specific fields due to the diversity of their possible meanings. To ensure the most rigorous literature search possible, no restrictions were placed on the publication date or article type during searching. 

Many AI approaches build on statistical models, such as logistic regression, which can blur the lines between disciplines. When conducting searches, a previously reported methodology was adopted \cite{Dhiman2022} whereby typical AI approaches were searched by name (e.g. neural networks), and other methods were searched by whether the authors described their work as \emph{artificial intelligence}. Full details of the search implementation for each database are provided in Appendix \ref{app:search}. The review protocol was registered with PROSPERO before the search results were screened for inclusion (CRD42022334730). 

\subsection*{Literature Selection}
One researcher (JB) manually removed duplicate papers with the assistance of the referencing software \emph{EndNote X9}. Two researchers (JB, KA) then independently screened articles for inclusion in two stages, the first based on title and abstract, the second based on full text. Disagreements were discussed and arbitrated by a third researcher (NR or NMO). Trials in WHO-ICTRP do not have associated abstracts, so for these studies, only titles were available for initial screening. 

The inclusion criteria required that research evaluated the use of at least one AI approach to make diagnostic or prognostic inferences on human histopathology images from suspected or confirmed cases of ovarian cancer. Studies were only included where AI methods were applied directly to the digital pathology images, or to features which were automatically extracted from the images. Fundamental tasks, such as segmentation and cell counting, were included as these could be used by pathologists for computer-aided diagnosis. Only conventional light microscopy images were considered, with other imaging modalities, such as fluorescence and hyperspectral imaging, excluded. Publications which did not include primary research were excluded (such as review papers). Non-English language articles and research where a full version of the manuscript was not accessible were excluded. 

\textcolor{black}{A model in an included study was considered to be a \emph{model of interest} if it met the same inclusion criteria. Where multiple models were compared against the same outcome, the model of interest was taken to be the newly proposed model, with the best performing model during validation taken if this was unclear. If multiple model outcomes were assessed in the same study, a model of interest was taken for each model outcome, regardless of any similarity in modelling approaches. The same model outcome at different levels of precision (e.g. patch-level, slide-level, patient-level) were not considered to be different model outcomes. Models didn't need to be entirely independent, for example, the output of one model of interest could have been used as the input of another model of interest on the condition that model performance was separately evaluated for each model.}

\subsection*{Risk of Bias \textcolor{black}{Assessment}}
The risk of bias \textcolor{black}{was assessed for models of interest} using the Prediction model Risk Of Bias ASsessment Tool (PROBAST) \citep{Wolff2019}, \textcolor{black}{where \emph{risk of bias} is the chance of reported results being distorted by limitations within the study design, conduct, and analysis. It includes 20 guiding questions which are} categorised into four domains (participants, predictors, outcome, and analysis), which are summarised as either high-risk or low-risk, or \textcolor{black}{unclear in the case that there is insufficient information to make a comprehensive assessment and none of the available information indicates a high risk of bias. As such, an unclear risk of bias does not indicate methodological flaws, but incomplete reporting. }

\textcolor{black}{The \textbf{participants} domain covers the recruitment and selection of participants to ensure the study population is consistent and representative of the target population. Relevant details include the participant recruitment strategy (when and where participants were recruited), the inclusion criteria, and how many participants were recruited.} 

\textcolor{black}{The \textbf{predictors} domain covers the consistent definition and measurement of predictors, which in this field typically refers to the generation of digital pathology images. This includes methods for fixing, staining, scanning, and digitally processing tissue before modelling. }

\textcolor{black}{The \textbf{outcome} domain covers the appropriate definition and consistent determination of ground-truth labels. This includes the criteria used to determine diagnosis/prognosis, the expertise of any persons determining these labels, and whether labels are determined independently of any model outputs.}

\textcolor{black}{The \textbf{analysis} domain covers statistical considerations in the evaluation of model performance to ensure valid and not unduly optimistic results. This includes many factors, such as the number of participants in the test set with each outcome, the validation approaches used (cross-validation, external validation, bootstrapping, etc.), the metrics used to assess performance, and methods used to overcome the effects of censoring, competing risks/confounders, and missing data. The risks caused by some of these factors are interrelated, for example, the risk of bias from using a small dataset is somewhat mitigated by cross-validation, which increases the effective size of the test set and can be used to assess variability, reducing optimism in the results. Further, the risk caused by using a small dataset depends on the type of outcome being predicted, for example, more data is required for a robust analysis of 5-class classification than binary classification. There must also be sufficient data within all relevant patient subgroups, for example, if multiple subtypes of ovarian cancer are included, there must not be a subtype that is only represented by a few patients. Due to these interrelated factors, there are no strict criteria to determine the appropriate size of a dataset, though fewer than 50 samples per class or fewer than 100 samples overall is likely to be considered high-risk, and more than 1000 samples overall is likely to be considered low-risk.} 

\textcolor{black}{Risks of bias often arise due to inconsistent methodologies. Inconsistency in the participants and predictors domains may cause heterogeneity in the visual properties of digital pathology slides which may lead to spurious correlations, either through random chance or systematic differences between subgroups in the dataset. Varied data may be beneficial during training to improve model generalisability when using large datasets, though this must be closely controlled to avoid introducing systematic confounding. Inconsistent determination of the outcome can mean that the results of a study are unreliable due to spurious correlations in the ground truth labels, or invalid due to incorrect determination of labels. }

\textcolor{black}{While PROBAST provides a framework to assess risks of bias, there is some level of subjectivity in the interpretation of signalling questions. As such}, each model was analysed by three independent researchers (any of JB, KA, NR, KZ, NMO), with at least one computer scientist and one clinician involved in the risk of bias assessment for each model. The PROBAST applicability of research analysis was not implemented as it is unsuitable for such a diverse array of possible research questions.

\subsection*{Data Synthesis}
Data extraction was performed independently by two researchers (JB, KA) using a form containing 81 fields within the categories \emph{Overview}, \emph{Data}, \emph{Methods}, \emph{Results}, and \emph{Miscellaneous}. Several of these fields were added or clarified during data extraction with the agreement of both researchers and retroactively applied to all accepted literature. The final data extraction form is available at \url{www.github.com/scjjb/OvCaReview}, with a summary included in Appendix~\ref{app:extract}.

Information was sought from full-text articles, as well as references and supplementary materials where appropriate. Inferences were made only when both researchers were confident that this gave the correct information, with disagreements resolved through discussion. Fields which could not be confidently completed were labelled as being \emph{unclear}.

All extracted data were summarised in two tables, one each for study-level and model-level characteristics. \textcolor{black}{Only models of interest were included in these tables. The term \emph{model outcome} refers to the model output, whether this was a clinical outcome (diagnosis/prognosis), or a diagnostically relevant outcome that could be used for computer-aided diagnosis, such as tumour segmentation.} The data synthesis did not include any meta-analysis due to the diversity of included methods and \textcolor{black}{model} outcomes.

\section*{Results}

\begin{figure}[h!]
\centering
\includegraphics[width=\textwidth]{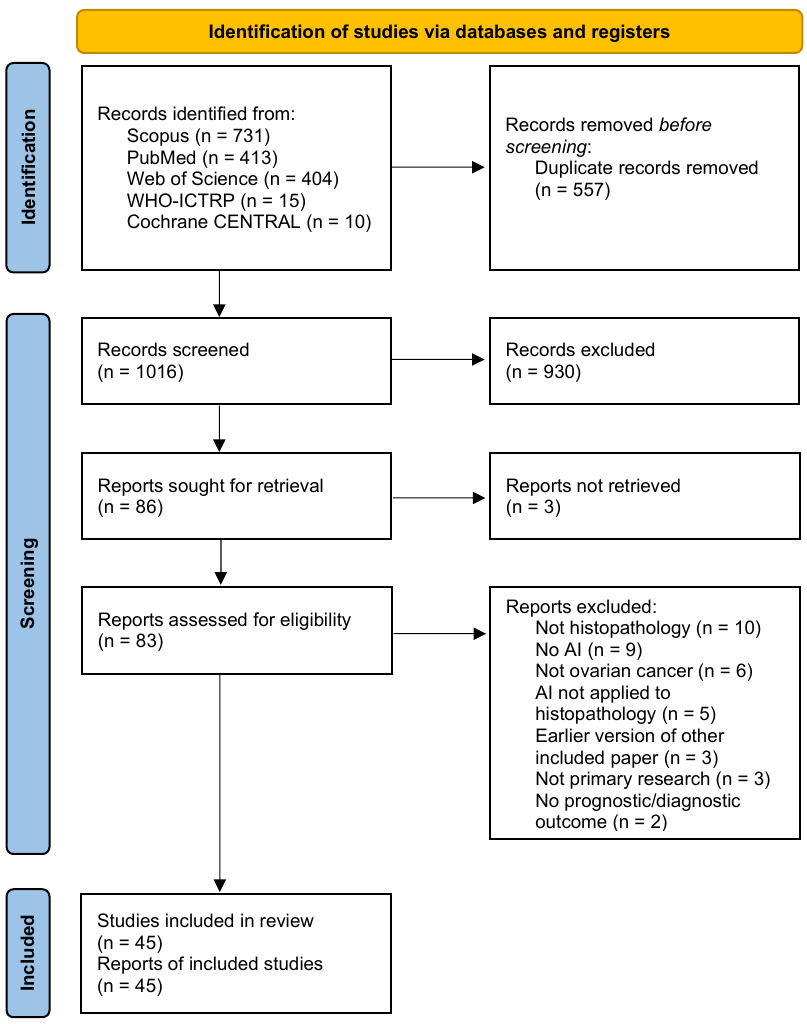}
\caption{PRISMA 2020 flowchart of the study identification and selection process for the systematic review. Records were screened on titles and abstracts alone, and reports were assessed based on the full-text content. \textcolor{black}{CENTRAL - Central Register of Controlled Trials. WHO-ICTRP - World Health Organisation International Clinical Trial Registry Platform.}}
\label{fig:flowchart}
\end{figure}

As shown in Figure \ref{fig:flowchart}, the literature searches returned a total of \textcolor{black}{1573} records, of which \textcolor{black}{557} were duplicates. \textcolor{black}{930} records were excluded during the screening of titles and abstracts, and \textcolor{black}{41} were excluded based on full paper screening, including \textcolor{black}{3} records for which full articles could not be obtained. The remaining \textcolor{black}{45} studies were included in the review, of which \textcolor{black}{11} were conference papers and \textcolor{black}{34} were journal papers. All accepted studies were originally identified through searches of research databases, with no records from trial registries meeting the inclusion criteria. While the searches returned literature from as early as 1949, all of the research which met the inclusion criteria was published since 2010, with \textcolor{black}{over 70\%} of the included literature published since 2020. Study characteristics are shown in Table \ref{table:characteristics}. The \textcolor{black}{45} accepted articles contained \textcolor{black}{80} models of interest, details of which are shown in Table \ref{table:models}.


\subsection*{Risk of Bias Assessment}

\begin{figure}[htbp]
\centering
\includegraphics[width=\textwidth]{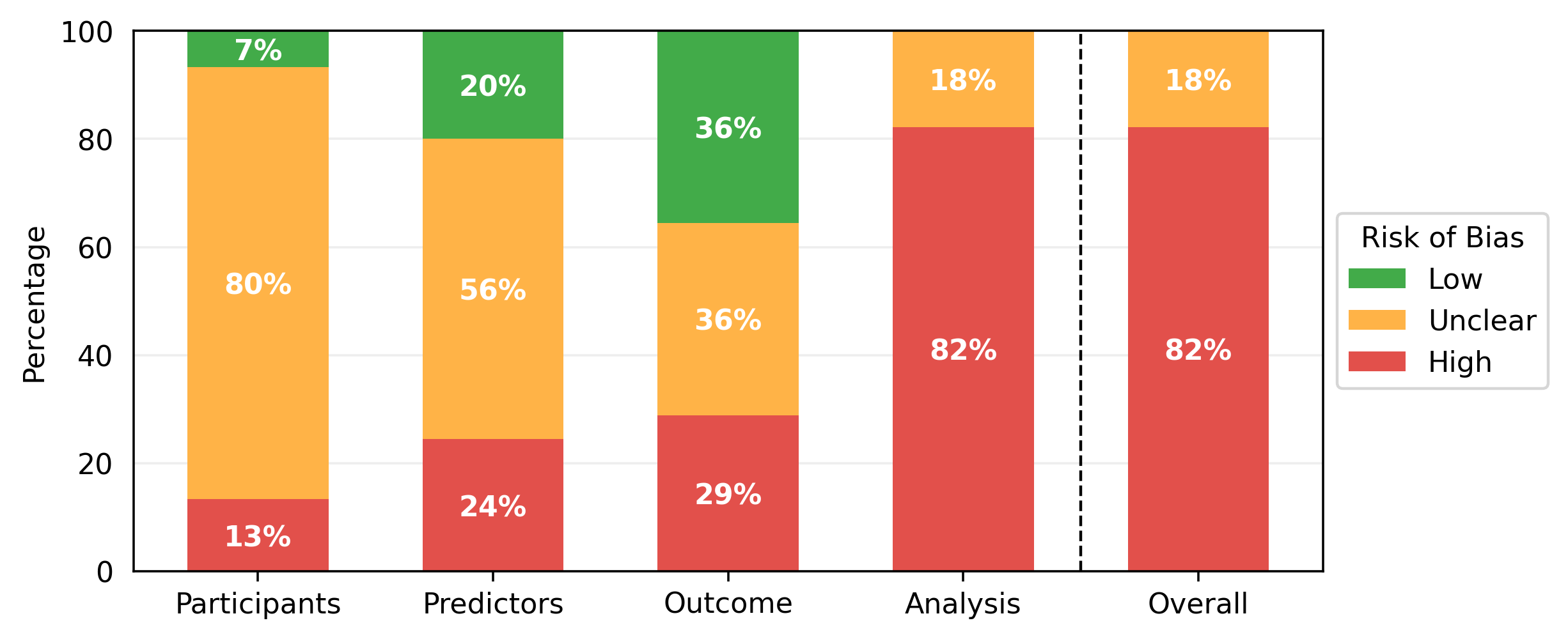}
\caption{PROBAST risk of bias results summarised for the \textcolor{black}{45} papers included in this review.}
\label{fig:probast}
\end{figure}
The results of the PROBAST assessments are shown in Table \ref{table:probast}. While some studies contained multiple models of interest, none of these contained models with different risk of bias scores for any section of the PROBAST assessment, so one risk of bias analysis is presented per paper.  All models showed either a high overall risk of bias (\textcolor{black}{37/45}) or an unclear overall risk of bias (\textcolor{black}{8/45}). Every high-risk model had a high-risk score in the analysis section (\textcolor{black}{37/45}), with several also being at high risk for participants (\textcolor{black}{6/45}), predictors (\textcolor{black}{11/45}), or outcomes (\textcolor{black}{13/45}). \textcolor{black}{Less than} half of the studies achieved a low risk of bias in any domain (\textcolor{black}{21/45}), with most low risks being found in the outcomes (\textcolor{black}{16/45}) and predictors (\textcolor{black}{9/45}) sections. Nearly all of the papers had an unclear risk of bias in at least one domain, most commonly the participants (\textcolor{black}{36/45}) and predictors (\textcolor{black}{25/45}) domains. Qualitative summaries are presented in Figure \ref{fig:probast}.

\begin{table}
\centering
\rowcolors{2}{gray!20}{gray!5}
 \begin{tabular}{|c|cccc|c|} 
 \hline
 \rowcolor{black!70}
\hspace{0.2cm}\color{white}\text{Publication}\hspace{0.2cm} & \color{white}\text{Participants} & \color{white}\text{Predictors} & \color{white}\text{Outcome} & \color{white}\text{Analysis} & \hspace{0.2cm}\color{white}\text{Overall}\hspace{0.2cm} \\
 \hline
 Dong 2010(a) \cite{Dong2010a} & \cellcolor{red!25}High & \cellcolor{red!25}High & \cellcolor{red!25}High & \cellcolor{red!25}High & \cellcolor{red!25}\textbf{High} \\ 
 Dong 2010(b) \cite{Dong2010b} & \cellcolor{red!25}High & \cellcolor{red!25}High & \cellcolor{red!25}High & \cellcolor{red!25}High & \cellcolor{red!25}\textbf{High} \\ 
 Signolle 2010 \cite{Signolle2010} & \cellcolor{yellow!25}Unclear & \cellcolor{yellow!25}Unclear & \cellcolor{red!25}High & \cellcolor{red!25}High & \cellcolor{red!25}\textbf{High} \\ 
 Janowczyk 2011 \cite{Janowczyk2011} & \cellcolor{yellow!25}Unclear & \cellcolor{yellow!25}Unclear & \cellcolor{green!25}Low & \cellcolor{red!25}High & \cellcolor{red!25}\textbf{High} \\ 
 Janowczyk 2012 \cite{Janowczyk2012} & \cellcolor{yellow!25}Unclear & \cellcolor{red!25}High & \cellcolor{yellow!25}Unclear & \cellcolor{red!25}High & \cellcolor{red!25}\textbf{High} \\ 
 Kothari 2012 \cite{Kothari2012} & \cellcolor{yellow!25}Unclear & \cellcolor{green!25}Low & \cellcolor{green!25}Low & \cellcolor{yellow!25}Unclear & \cellcolor{yellow!25}\textbf{Unclear} \\ 
 Poruthoor 2013 \cite{Poruthoor2013} & \cellcolor{yellow!25}Unclear & \cellcolor{red!25}High & \cellcolor{red!25}High & \cellcolor{red!25}High & \cellcolor{red!25}\textbf{High} \\ 
 BenTaieb 2015 \cite{BenTaieb2015} & \cellcolor{yellow!25}Unclear & \cellcolor{yellow!25}Unclear & \cellcolor{green!25}Low & \cellcolor{red!25}High & \cellcolor{red!25}\textbf{High} \\ 
 BenTaieb 2016 \cite{BenTaieb2016} & \cellcolor{yellow!25}Unclear & \cellcolor{red!25}High & \cellcolor{yellow!25}Unclear & \cellcolor{red!25}High & \cellcolor{red!25}\textbf{High} \\ 
 BenTaieb 2017 \cite{BenTaieb2017} & \cellcolor{yellow!25}Unclear & \cellcolor{yellow!25}Unclear & \cellcolor{green!25}Low & \cellcolor{red!25}High & \cellcolor{red!25}\textbf{High} \\ 
 Lorsakul 2017 \cite{Lorsakul2017} & \cellcolor{yellow!25}Unclear & \cellcolor{yellow!25}Unclear & \cellcolor{red!25}High & \cellcolor{red!25}High & \cellcolor{red!25}\textbf{High} \\ 
 Du 2018 \cite{Du2018} & \cellcolor{yellow!25}\hspace{0.25cm}Unclear\hspace{0.25cm} & \cellcolor{yellow!25}\hspace{0.25cm}Unclear\hspace{0.25cm} & \cellcolor{yellow!25}\hspace{0.25cm}Unclear\hspace{0.25cm} & \cellcolor{yellow!25}\hspace{0.25cm}Unclear\hspace{0.25cm} & \cellcolor{yellow!25}\hspace{0.25cm}\textbf{Unclear}\hspace{0.25cm} \\ 
 Heindl 2018 \cite{Heindl2018} & \cellcolor{yellow!25}Unclear & \cellcolor{green!25}Low & \cellcolor{green!25}Low & \cellcolor{red!25}High & \cellcolor{red!25}\textbf{High} \\ 
 Kalra 2020 \cite{Kalra2020} & \cellcolor{yellow!25}Unclear & \cellcolor{green!25}Low & \cellcolor{green!25}Low & \cellcolor{red!25}High & \cellcolor{red!25}\textbf{High} \\ 
 Levine 2020 \cite{Levine2020} & \cellcolor{yellow!25}Unclear & \cellcolor{green!25}Low & \cellcolor{green!25}Low & \cellcolor{yellow!25}Unclear & \cellcolor{yellow!25}\textbf{Unclear} \\ 
 Yaar 2020 \cite{Yaar2020} & \cellcolor{yellow!25}Unclear & \cellcolor{yellow!25}Unclear & \cellcolor{green!25}Low & \cellcolor{red!25}High & \cellcolor{red!25}\textbf{High} \\ 
 Yu 2020 \cite{Yu2020} & \cellcolor{yellow!25}Unclear & \cellcolor{green!25}Low & \cellcolor{green!25}Low & \cellcolor{red!25}High & \cellcolor{red!25}\textbf{High} \\ 
 Gentles 2021 \cite{Gentles2021} & \cellcolor{red!25}High & \cellcolor{yellow!25}Unclear & \cellcolor{red!25}High & \cellcolor{red!25}High & \cellcolor{red!25}\textbf{High} \\ 
 Ghoniem 2021 \cite{Ghoniem2021} & \cellcolor{yellow!25}Unclear & \cellcolor{yellow!25}Unclear & \cellcolor{yellow!25}Unclear & \cellcolor{red!25}High & \cellcolor{red!25}\textbf{High} \\ 
 Jiang 2021 \cite{Jiang2021} & \cellcolor{red!25}High & \cellcolor{red!25}High & \cellcolor{yellow!25}Unclear & \cellcolor{red!25}High & \cellcolor{red!25}\textbf{High} \\ 
 Laury 2021 \cite{Laury2021} & \cellcolor{green!25}Low & \cellcolor{red!25}High & \cellcolor{red!25}High & \cellcolor{red!25}High & \cellcolor{red!25}\textbf{High} \\ 
 Paijens 2021 \cite{Paijens2021} & \cellcolor{green!25}Low & \cellcolor{red!25}High & \cellcolor{yellow!25}Unclear & \cellcolor{red!25}High & \cellcolor{red!25}\textbf{High} \\ 
 Shin 2021 \cite{Shin2021} & \cellcolor{yellow!25}Unclear & \cellcolor{yellow!25}Unclear & \cellcolor{yellow!25}Unclear & \cellcolor{red!25}High & \cellcolor{red!25}\textbf{High} \\ 
 Zeng 2021 \cite{Zeng2021} & \cellcolor{yellow!25}Unclear & \cellcolor{yellow!25}Unclear & \cellcolor{green!25}Low & \cellcolor{red!25}High & \cellcolor{red!25}\textbf{High} \\ 
 Boehm 2022 \cite{Boehm2022} & \cellcolor{yellow!25}Unclear & \cellcolor{red!25}High & \cellcolor{yellow!25}Unclear & \cellcolor{red!25}High & \cellcolor{red!25}\textbf{High} \\ 
 Boschman 2022 \cite{Boschman2022} & \cellcolor{yellow!25}Unclear & \cellcolor{green!25}Low & \cellcolor{green!25}Low & \cellcolor{red!25}High & \cellcolor{red!25}\textbf{High} \\ 
 Elie 2022 \cite{Elie2022} & \cellcolor{yellow!25}Unclear & \cellcolor{green!25}Low & \cellcolor{red!25}High & \cellcolor{red!25}High & \cellcolor{red!25}\textbf{High} \\ 
 Farahani 2022 \cite{Farahani2022} & \cellcolor{yellow!25}Unclear & \cellcolor{yellow!25}Unclear & \cellcolor{green!25}Low & \cellcolor{yellow!25}Unclear & \cellcolor{yellow!25}\textbf{Unclear} \\ 
 Hu 2022 \cite{Hu2022} & \cellcolor{yellow!25}Unclear & \cellcolor{yellow!25}Unclear & \cellcolor{yellow!25}Unclear & \cellcolor{yellow!25}Unclear & \cellcolor{yellow!25}\textbf{Unclear} \\ 
 Jiang 2022 \cite{Jiang2022} & \cellcolor{yellow!25}Unclear & \cellcolor{yellow!25}Unclear & \cellcolor{red!25}High & \cellcolor{red!25}High & \cellcolor{red!25}\textbf{High} \\ 
 Kasture 2022 \cite{Kasture2022} & \cellcolor{red!25}High & \cellcolor{red!25}High & \cellcolor{red!25}High & \cellcolor{red!25}High & \cellcolor{red!25}\textbf{High} \\ 
 Kowalski 2022 \cite{Kowalski2022} & \cellcolor{yellow!25}Unclear & \cellcolor{yellow!25}Unclear & \cellcolor{yellow!25}Unclear & \cellcolor{red!25}High & \cellcolor{red!25}\textbf{High} \\ 
 \textcolor{black}{Lazard 2022 \cite{Lazard2022}} & \cellcolor{yellow!25}Unclear & \cellcolor{yellow!25}Unclear & \cellcolor{yellow!25}Unclear & \cellcolor{yellow!25}Unclear & \cellcolor{yellow!25}\textbf{Unclear} \\ 
 Liu 2022 \cite{Liu2022} & \cellcolor{yellow!25}Unclear & \cellcolor{yellow!25}Unclear & \cellcolor{yellow!25}Unclear & \cellcolor{yellow!25}Unclear & \cellcolor{yellow!25}\textbf{Unclear} \\ 
\textcolor{black}{ Mayer 2022 \cite{Mayer2022}} & \cellcolor{yellow!25}Unclear & \cellcolor{yellow!25}Unclear & \cellcolor{red!25}High & \cellcolor{red!25}High & \cellcolor{red!25}\textbf{High} \\ 
 Nero 2022 \cite{Nero2022} & \cellcolor{yellow!25}Unclear & \cellcolor{green!25}Low & \cellcolor{red!25}High & \cellcolor{red!25}High & \cellcolor{red!25}\textbf{High} \\ 
 Salguero 2022 \cite{Salguero2022} & \cellcolor{yellow!25}Unclear & \cellcolor{yellow!25}Unclear & \cellcolor{green!25}Low & \cellcolor{red!25}High & \cellcolor{red!25}\textbf{High} \\ 
 \textcolor{black}{Wang 2022(a) \cite{Wang2022a}} & \cellcolor{yellow!25}Unclear & \cellcolor{yellow!25}Unclear & \cellcolor{yellow!25}Unclear & \cellcolor{red!25}High & \cellcolor{red!25}\textbf{High} \\ 
 Wang 2022\textcolor{black}{(b)} \cite{Wang2022b} & \cellcolor{yellow!25}Unclear & \cellcolor{yellow!25}Unclear & \cellcolor{green!25}Low & \cellcolor{red!25}High & \cellcolor{red!25}\textbf{High} \\ 
 \textcolor{black}{Yokomizo 2022 \cite{Yokomizo2022}} & \cellcolor{green!25}Low & \cellcolor{green!25}Low & \cellcolor{yellow!25}Unclear & \cellcolor{yellow!25}Unclear & \cellcolor{yellow!25}\textbf{Unclear} \\ 
 \textcolor{black}{Ho 2023 \cite{Ho2023}} & \cellcolor{yellow!25}Unclear & \cellcolor{yellow!25}Unclear & \cellcolor{yellow!25}Unclear & \cellcolor{red!25}High & \cellcolor{red!25}\textbf{High} \\ 
 \textcolor{black}{Meng 2023 \cite{Meng2023}} & \cellcolor{yellow!25}Unclear & \cellcolor{yellow!25}Unclear & \cellcolor{green!25}Low & \cellcolor{red!25}High & \cellcolor{red!25}\textbf{High} \\ 
 \textcolor{black}{Ramasamy 2023 \cite{Ramasamy2023}} & \cellcolor{red!25}High & \cellcolor{red!25}High & \cellcolor{red!25}High & \cellcolor{red!25}High & \cellcolor{red!25}\textbf{High} \\ 
 \textcolor{black}{Wang 2023 \cite{Wang2023}} & \cellcolor{yellow!25}Unclear & \cellcolor{yellow!25}Unclear & \cellcolor{yellow!25}Unclear & \cellcolor{red!25}High & \cellcolor{red!25}\textbf{High} \\ 
 \textcolor{black}{Wu 2023 \cite{Wu2023}} & \cellcolor{yellow!25}Unclear & \cellcolor{yellow!25}Unclear & \cellcolor{green!25}Low & \cellcolor{red!25}High & \cellcolor{red!25}\textbf{High} \\ 
 \hline
 \end{tabular}
 \caption{PROBAST risk of bias assessment results for the \textcolor{black}{45} papers included in this review. This is presented as one row for each paper because every paper that contained multiple models of interest was found to have the same risk of bias for every model.}
 \label{table:probast}
\end{table}

\newpage
\subsection*{Data Synthesis Results}

\subsubsection*{Data in Included Literature}
The number of participants in internal datasets varied by orders of magnitude, with each study including 1 to \textcolor{black}{776} ovarian cancer patients, and one study including over 10,000 total patients across a range of 32 malignancies \cite{Kalra2020}.
\textcolor{black}{Most research only used data from the five most common subtypes of ovarian carcinoma, though one recent study included the use of sex cord-stromal tumours \cite{Meng2023}}. 
Only one study explicitly included any prospective data collection, and this was only for a small subset which was not used for external validation \cite{Boehm2022}. 

\begin{figure}[htbp]
    \centering
    \begin{subfigure}{.5\textwidth}
    \centering
    \includegraphics[trim={0.4cm 0.4cm 0.4cm 0.4cm },width=\linewidth]{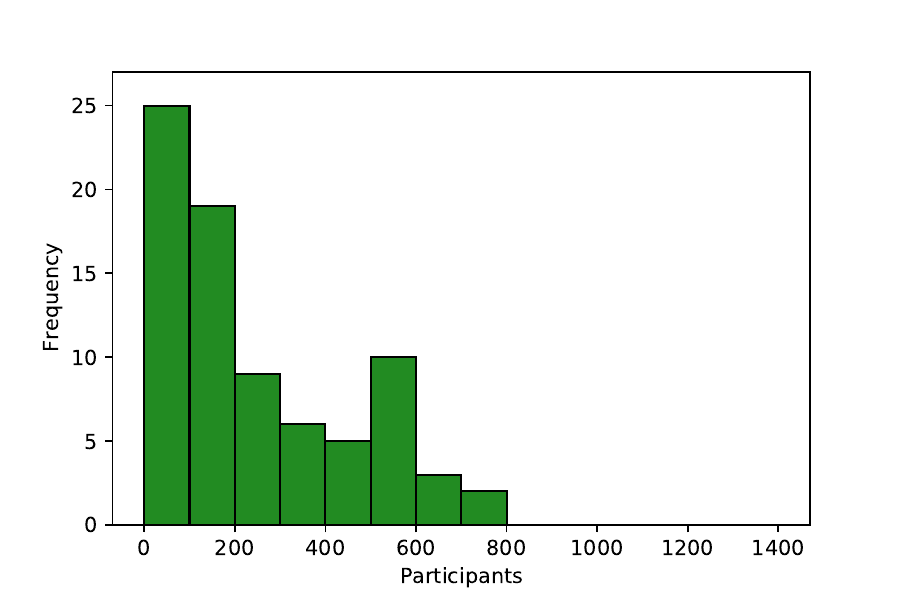}
    \end{subfigure}%
    \begin{subfigure}{.5\textwidth}
    \centering
    \includegraphics[trim={0.4cm 0.4cm 0.4cm 0.4cm},width=\linewidth]{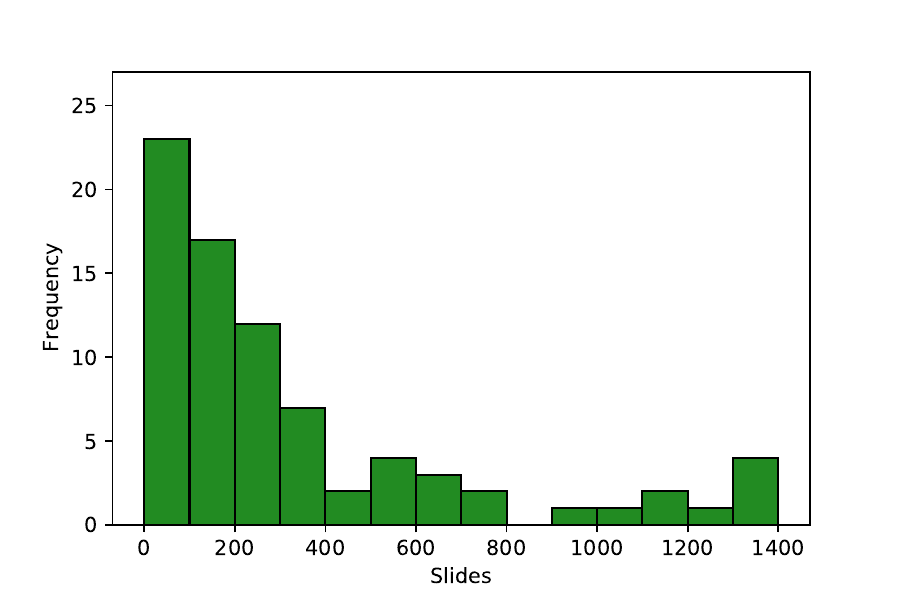}
    \end{subfigure}%
    \caption{Histograms showing the number of ovarian cancer patients and slides used in model development. Many of these values are uncertain due to incomplete reporting, as reflected in Table \ref{table:models}.}
    \label{fig:slides}
\end{figure}

As shown in Figure \ref{fig:slides}, the number of pathology slides used was often much greater than the number of patients included, with three studies using over 1000 slides from ovarian cancer patients \cite{Kothari2012, Yu2020, Liu2022}. In most of the studies, \textcolor{black}{model development samples were WSIs containing resected or biopsied tissue} (34/45), with others using \textcolor{black}{individual} tissue microarray (TMA) \textcolor{black}{core images}  (\textcolor{black}{5/45}) or pre-cropped digital pathology images (\textcolor{black}{3/45}). Most studies used H\&E-stained tissue (\textcolor{black}{33/45}) and others used a variety of IHC stains (\textcolor{black}{11/45}), with no two papers reporting the use of the same IHC stains. Some studies included multi-modal approaches, using genomics 
\cite{Poruthoor2013,Yaar2020,Ghoniem2021,Zeng2021,Boehm2022}, proteomics \cite{Poruthoor2013,Zeng2021}, transcriptomics \cite{Zeng2021}, and radiomics \cite{Boehm2022} data alongside histopathological data.

The most commonly used data source was The Cancer Genome Atlas (TCGA) (\textcolor{black}{18/45}),
a project from which over 30,000 digital pathology images from 33 malignancies are publicly available. The ovarian cancer subset, TCGA-OV \cite{Holback2016}, contains 1481 WSIs from 590 cases of ovarian serous carcinoma (mostly, but not exclusively, high-grade), with corresponding genomic, transcriptomic, and clinical data. This includes slides from eight data centres in the United States, with most slides containing frozen tissue sections (1374/1481) rather than formalin-fixed, paraffin-embedded (FFPE) sections. Other recurring data sources were the University of British Columbia Ovarian Cancer Research Program (OVCARE) repository \cite{Levine2020,Boschman2022,Farahani2022}, the Transcanadian study \cite{BenTaieb2015,BenTaieb2016}, \textcolor{black}{and clinical records at the Mayo Clinic \cite{Jiang2021,Jiang2022}, Tri-Service General Hospital \cite{Wang2022a,Wang2022b,Wang2023}, and Memorial Sloan
Kettering Cancer Center \cite{Boehm2022,Ho2023}}. All other researchers either used a unique data source (\textcolor{black}{12/45}) or did not report the provenance of their data (8/\textcolor{black}{45}). TCGA-OV, OVCARE, and the Transcanadian study are all multi-centre datasets. Aside from these, few studies reported the use of multi-centre data \textcolor{black}{\cite{Paijens2021, Shin2021, Zeng2021, Boehm2022, Farahani2022, Mayer2022}}. Only two studies reported the use of multiple slide scanners, with every slide scanned on one of two available scanners \cite{Boschman2022, Farahani2022}. The countries from which data were sourced included Canada, China, Finland, France, \textcolor{black}{Germany,} Italy, \textcolor{black}{Japan,} the Netherlands, South Korea, Taiwan, the United Kingdom, and the United States of America.

\subsubsection*{Methods in Included Literature}

There was a total of \textcolor{black}{80} models of interest in the \textcolor{black}{45} included papers, with each paper containing 1 to 6 such models. \textcolor{black}{There were 37 diagnostic models, 22 prognostic models, and 21 other models predicting diagnostically relevant information. Diagnostic model outcomes included the classification of malignancy status (10/37), histological subtype (7/37), primary cancer type (5/37), genetic mutation status (4/37), tumour-stroma reaction level (3/37), grade (2/37), transcriptomic subtype (2/37), stage (1/37), microsatellite instability status (1/37), epithelial-mesenchymal transition status (1/37), and homologous recombination deficiency status (1/37). Prognostic models included the prediction of treatment response (11/23), overall survival (6/23), progression-free survival (3/23), and recurrence (2/23). The other models performed tasks which could be used to assist pathologists in analysing pathology images, including measuring the quantity/intensity of staining, generating segmentation masks, and classifying tissue/cell types.} 


A variety of models were used, with the most common types being convolutional neural network (CNN) (\textcolor{black}{41/80}), support vector machine (SVM) (10/\textcolor{black}{80}), and random forest (\textcolor{black}{6/80}). CNN architectures included GoogLeNet \cite{Du2018}, \textcolor{black}{VGG16 \cite{Yu2020, Jiang2022}}, VGG19 \cite{Levine2020, Farahani2022}, InceptionV3 \textcolor{black}{\cite{Shin2021, Wang2022a, Wang2022b, Wang2023}}, ResNet18 \textcolor{black}{\cite{Boehm2022, Boschman2022, Farahani2022, Hu2022, Lazard2022, Mayer2022}, ResNet34 \cite{Yokomizo2022},} ResNet50 \textcolor{black}{\cite{Nero2022, Meng2023, Wu2023}, ResNet182 \cite{Ho2023}}, and MaskRCNN \cite{Jiang2022}. Novel CNNs typically used multiple standardised blocks involving convolutional, normalization, activation, and/or pooling layers \cite{Yaar2020, Kasture2022, Kowalski2022}, with \textcolor{black}{two studies} also including attention modules \textcolor{black}{\cite{Liu2022, Wang2023}}. One study generated their novel architecture by using a topology optimization approach on a standard VGG16 \cite{Ghoniem2021}. 

Most researchers split their original images into patches to be separately processed, with patch sizes ranging from 60x60 to 2048x2048 pixels, the most common being 512x512 pixels (\textcolor{black}{19/56}) and 256x256 pixels (\textcolor{black}{12/56}). A range of feature extraction techniques were employed, including both hand-crafted/pre-defined features (\textcolor{black}{23/80}) and features that were automatically learned by the model (\textcolor{black}{51/80}). Hand-crafted features included a plethora of textural, chromatic, and cellular and nuclear morphological features. Hand-crafted features were commonly used as inputs to classical ML methods, such as SVM and random forest models. Learned features were typically extracted using a CNN, which was often also used for classification. 

Despite the common use of patches, most models made predictions at the WSI level (\textcolor{black}{29/80}), \textcolor{black}{TMA core level (18/80)}, or patient level (6/\textcolor{black}{80}), requiring aggregation of patch-level information. Two distinct aggregation approaches were used, one aggregating before modelling and one aggregating after modelling. The former approach requires the generation of slide-level features before modelling, the latter requires the aggregation of patch-level model outputs to make slide-level predictions. Slide-level features were generated using \textcolor{black}{summation \cite{Meng2023}}, averaging \textcolor{black}{\cite{Poruthoor2013, Zeng2021, Ho2023}}, attention-based weighted averaging \textcolor{black}{\cite{Hu2022, Liu2022, Nero2022, Lazard2022, Wu2023}}, concatenation \cite{BenTaieb2016, Kalra2020}, as well as more complex embedding approaches using Fisher vector encoding \cite{BenTaieb2015} and k-means clustering \cite{BenTaieb2017}.
Patch-level model outputs were aggregated to generate slide-level predictions by taking the maximum \textcolor{black}{\cite{Yaar2020, Wang2023}, median \cite{Yokomizo2022}}, or average \cite{Ghoniem2021}, using voting strategies \cite{Boschman2022, Wang2022b}, or using a random forest classifier \cite{Farahani2022}. These approaches are all examples of \emph{multiple instance learning} (MIL), though few models of interest were reported using this terminology \textcolor{black}{\cite{Yaar2020,Hu2022,Lazard2022,Nero2022}}.

\begin{table}
    \centering
    \includegraphics[trim={4cm 6.5cm 4cm 2.5cm},width=0.85\textwidth]{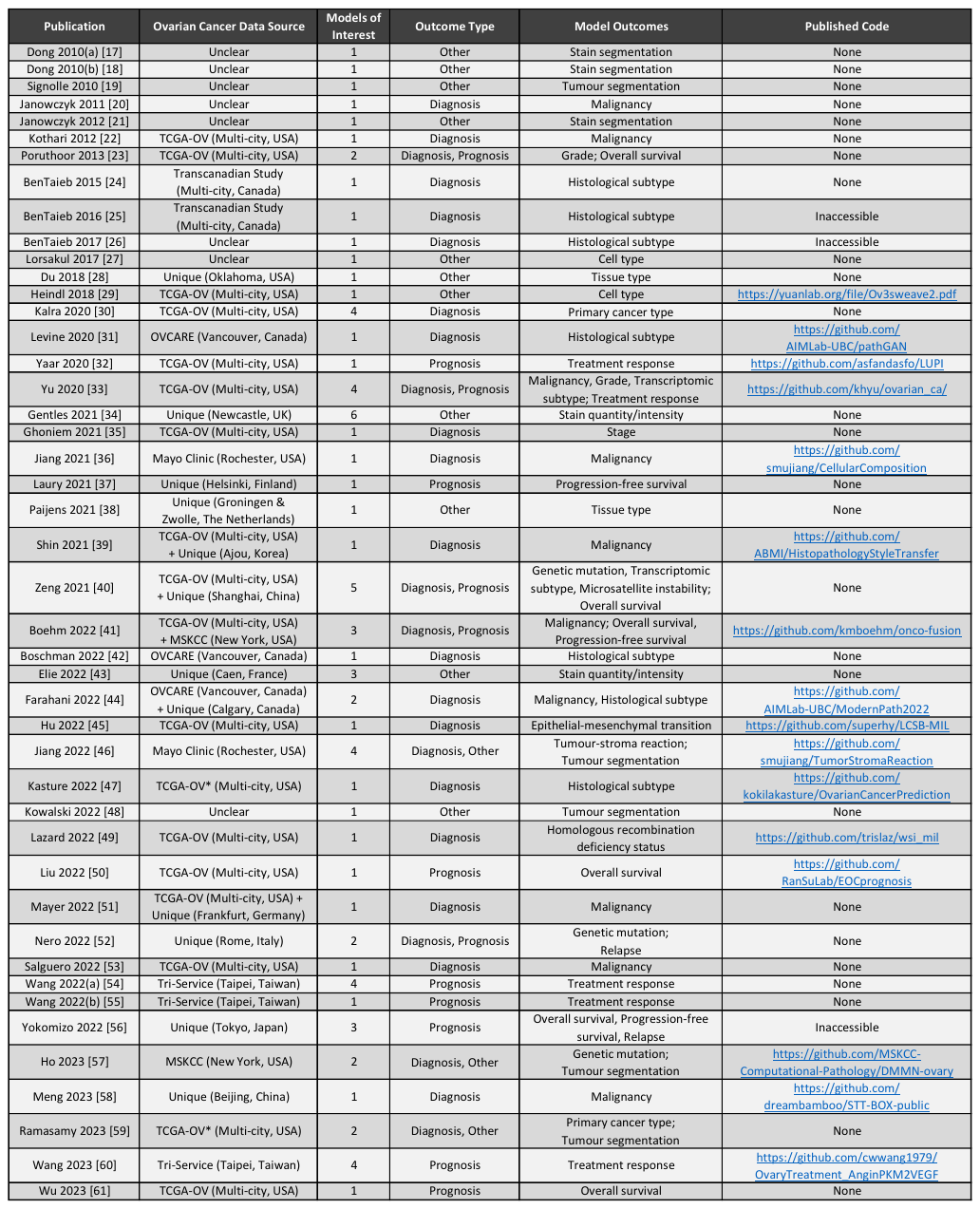}
\caption{Characteristics of the \textcolor{black}{45} studies included in this systematic review. Details are shown for individual models in Table \ref{table:models}. \textcolor{black}{Six} data sources are used in multiple studies - The Cancer Genome Atlas (TCGA-OV) \cite{Holback2016}, the British Columbia Ovarian Cancer Research Program (OVCARE), The Transcanadian Study \cite{Kobel2010}, \textcolor{black}{and three individual centres (Mayo Clinic, Tri-Service, and Memorial Sloan Kettering Cancer Center (MSKCC))}. Code is labelled as inaccessible where it could not be found despite a link being provided in the publication. \textcolor{black}{*Indicates papers where significant discrepancies were found regarding the data source, as described in the Discussion.}}
 \label{table:characteristics}
\end{table}

\newpage

\begin{landscape}
\thispagestyle{empty}
\begin{table}
    \centering
    \makebox[0pt]{
    \includegraphics[trim={1cm 0.15cm 1cm 2.5cm},width=23.7cm]{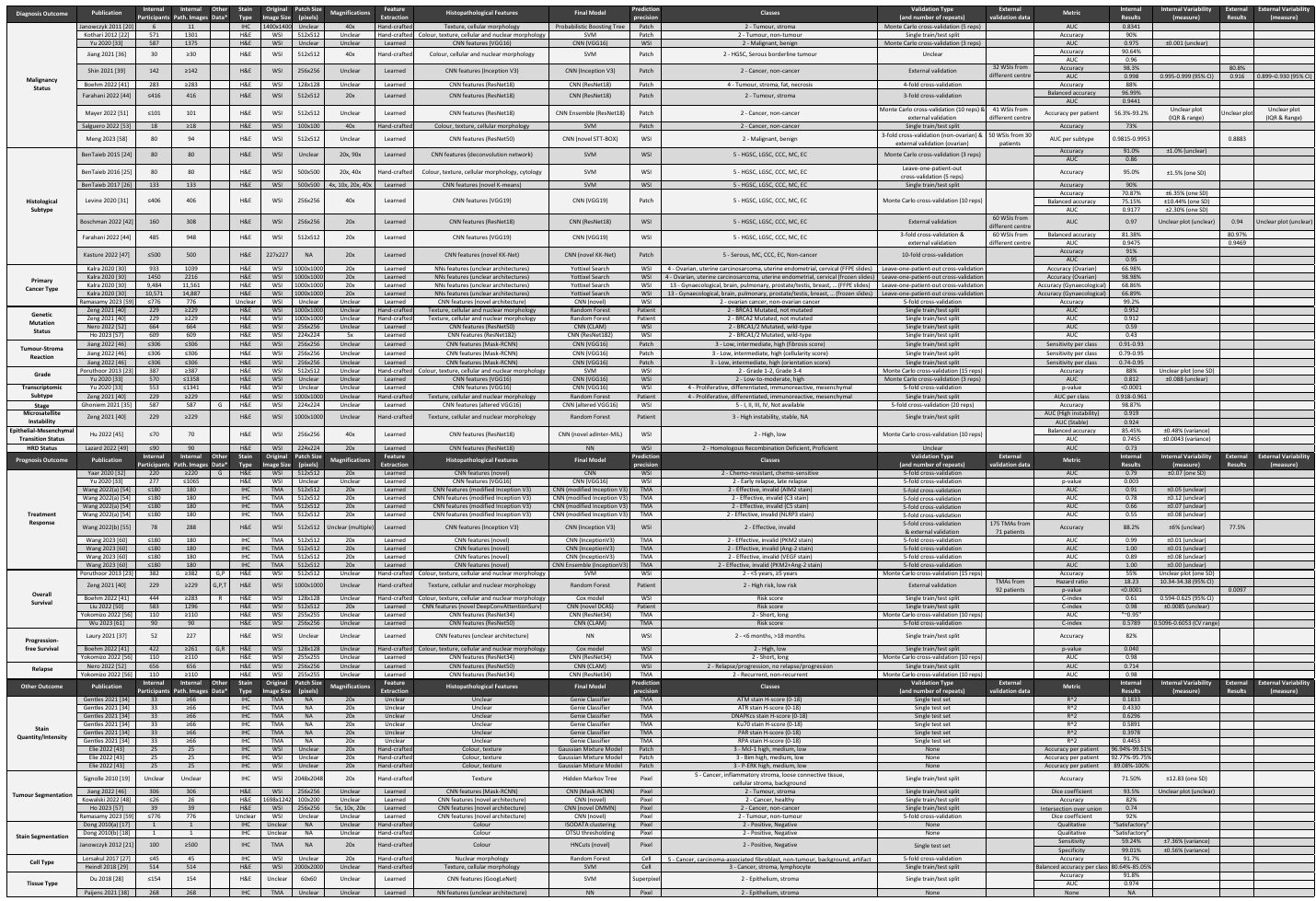}
    }
    \caption{Characteristics of the \textcolor{black}{80} models of interest from the \textcolor{black}{45} papers included in this systematic review, grouped by \textcolor{black}{model} outcome. *Other data types are Genomics (G), Proteomics (P), Radiomics (R), and Transcriptomics (T). SVM - support vector machine. CNN - convolutional neural network. AUC - area under the receiver operating characteristic (ROC) curve. HGSC - high-grade serous carcinoma. LGSC - low-grade serous carcinoma. CCC - clear cell carcinoma. MC - mucinous carcinoma. EC - endometrioid carcinoma. \textcolor{black}{H\&E - haematoxylin and eosin. IHC - immunohistochemistry. TMA refers to individual cores from tissue microarrays, WSI refers to whole slide images of biopsy or resection specimens.}}
    \label{table:models}
\end{table}

\end{landscape}

Most studies included segmentation at some stage, with many of these analysing tumour/stain segmentation as a model outcome \cite{Dong2010a, Dong2010b, Signolle2010, Janowczyk2011, Janowczyk2012, Paijens2021, Jiang2022, Kowalski2022, Ho2023, Ramasamy2023}. Some other studies used segmentation to determine regions of interest for further modelling, either simply separating tissue from background \cite{Kothari2012, Kalra2020, Nero2022, Wu2023}, or using tumour segmentation to select the most relevant tissue regions \cite{Gentles2021,Laury2021,Wang2022a,Wang2022b,Wang2023}. One study also used segmentation to detect individual cells for classification \cite{Heindl2018}. Some studies also used segmentation in determining hand-crafted features relating to the quantity and morphology of different tissues, cells, and nuclei \cite{Kothari2012, Poruthoor2013, BenTaieb2016, Jiang2021, Zeng2021, Boehm2022}.


While attention-based approaches have been applied to other malignancies for several years \cite{Ilse2018, Lu2021}, they were only seen in the most recent ovarian cancer studies \textcolor{black}{\cite{Farahani2022, Hu2022, Lazard2022, Liu2022, Nero2022,Wang2022a, Wang2022b, Wu2023, Wang2023}}, and none of the methods included self-attention, an increasingly popular method for other malignancies \cite{He2022}. Most models were deterministic, though hidden Markov trees \cite{Signolle2010}, probabilistic boosting trees \cite{Janowczyk2011}, and Gaussian mixture models \cite{Elie2022} were also used. \textcolor{black}{Aside from the common use of low-resolution images to detect and remove non-tissue areas, images were typically analysed at a single resolution, with only six papers including multi-magnification techniques in their models of interest. Four of these combined features from different resolutions for modelling \cite{BenTaieb2015, BenTaieb2016, BenTaieb2017, Ho2023}, and the other two used different magnifications for selecting informative tissue regions and for modelling \cite{Wang2022a, Wang2022b}}. 
Out of the papers for which it could be determined, the most common modelling magnifications were 20x (\textcolor{black}{35/41}) and 40x (7/\textcolor{black}{41}). Few models integrated histopathology data with other modalities (6/\textcolor{black}{80}). Multi-modal approaches included the concatenation of separately extracted uni-modal features before modelling \cite{Poruthoor2013, Ghoniem2021, Zeng2021}, the amalgamation of uni-modal predictions from separate models \cite{Boehm2022}, and a teacher-student approach where multiple modalities were used in model training but only histopathology data was used for prediction \cite{Yaar2020}.

\subsubsection*{Analysis in Included Literature}

Analyses were limited, with less than half of the \textcolor{black}{model} outcomes being evaluated with cross-validation \textcolor{black}{(39/80) and with very few externally validated using independent ovarian cancer data (7/80)}, despite small internal cohort sizes. Cross-validation methods included k-fold (\textcolor{black}{22/39}) with \textcolor{black}{3} to 10 folds, Monte Carlo (\textcolor{black}{12/39}) with 3 to 15 repeats, and leave-one-patient-out cross-validations (5/\textcolor{black}{39}). Some other papers included cross-validation on the training set to select hyperparameters but used only a small unseen test set from the same data source for evaluation. Externally validated models were all trained with WSIs, with validations either performed on TMA \textcolor{black}{cores} (\textcolor{black}{2/7}) or WSIs from independent data sources (\textcolor{black}{5/7}), with two of these explicitly using different scanners to digitize internal and external data \cite{Boschman2022, Farahani2022}.
Some reported methods were externally validated with data from non-ovarian malignancies, but none of these included ovarian cancer data in any capacity, so were not included in the review. \textcolor{black}{However, there was one method which trained with only gastrointestinal tumour data and externally validated with ovarian tumour data \cite{Meng2023}}.


Most classification models were evaluated using accuracy, balanced accuracy, and/or area under the receiver operating characteristic curve (AUC), with one exception where only a p-value was reported measuring the association between histological features and transcriptomic subtypes based on a Kruskal-Wallis test \cite{Yu2020}. Some models were also evaluated using the F1-score, which we chose not to tabulate (in Figure \ref{table:models}) as the other metrics were reported more consistently.
Survival model performance was typically reported using AUC, with other metrics including p-value, accuracy, hazard ratios, and \textcolor{black}{C-index, which is similar to AUC but can account for censoring}. Segmentation models were almost all evaluated differently from each other, with different studies reporting AUC, accuracy, Dice coefficient, \textcolor{black}{intersection over union}, sensitivity, specificity, and qualitative evaluations. Regression models were all evaluated using the coefficient of determination ($R^2$-statistic). \textcolor{black}{For some models, performance was broken down per patient \cite{Elie2022, Mayer2022}, per subtype \cite{Meng2023}, or per class \cite{Heindl2018,Kalra2020, Zeng2021, Jiang2022}, without an aggregated, holistic measure of model performance.}

The variability of model performance was not frequently reported (\textcolor{black}{33/94}), and when it was reported it was often incomplete. This included cases where it was unclear what the intervals represented (95\% confidence interval, one standard deviation, variation, etc.), or not clear what the exact bounds of the interval were due to results being plotted but not explicitly stated. Within the entire review, there were only \textcolor{black}{three} examples in which variability was reported during external validation \textcolor{black}{\cite{Shin2021, Boschman2022, Mayer2022}, only one of which clearly reported both the bounds and the type of the interval \cite{Shin2021}}. No studies performed any Bayesian form of uncertainty quantification. Reported results are shown in Table \ref{table:models}, though direct comparisons between the performance of different models should be treated with caution due to the diversity of data and validation methods used to evaluate different models, the lack of variability measures, the consistently high risks of bias, and the heterogeneity in reported metrics.



\section*{Discussion}

The vast majority of published research on AI for diagnostic or prognostic purposes in ovarian cancer histopathology was found to be at a high risk of bias due to issues within the analyses performed. Researchers often used a limited quantity of data and conducted analyses on a single train-test data split without using any methods to account for overfitting and model optimism (cross-validation, bootstrapping, external validation). \textcolor{black}{These limitations are common in gynaecological AI research using other data types, with recent reviews pointing to poor clinical utility caused by predominantly retrospective studies using limited data \cite{Shrestha2022, Zhou2022} and limited methodologies with weak validation, which risk model performance being overestimated \cite{Fiste2022, Xu2022}.} 

The more robust analyses included one study in which several relevant metrics were evaluated using 10 repeats of Monte Carlo cross-validation on a set of 406 WSIs, with standard deviations reported for each metric \cite{Levine2020}. Other positive examples included the use of both internal cross-validation and external validation for the same outcome, giving a more rigorous analysis \textcolor{black}{\cite{Farahani2022, Mayer2022, Wang2022b}}. While external validations were uncommon, those which were conducted offered a real insight into model generalisability, with a clear reduction in performance on all external validation sets except one \cite{Farahani2022}. The only study which demonstrated high generalisability included the largest training set out of all externally validated approaches, included more extensive data labelling than many similar studies, and implemented a combination of three colour normalisation approaches, indicating that these factors may benefit generalisability. 


Studies frequently had an unclear risk of bias within the participants and predictors domains of PROBAST \textcolor{black}{due to incomplete reporting. Frequently missing information included where the patients were recruited, how many patients were included, how many samples/images were used, whether any patients/images were excluded, and the methods by which tissue was processed and digitized. Reporting was often poor regarding open-access datasets.} Only \textcolor{black}{three} papers were found to be at low risk of bias for participants, with these including clear and reasonable patient recruitment strategies and selection criteria, which can be seen as positive examples for other researchers \textcolor{black}{\cite{Laury2021, Paijens2021, Yokomizo2022}}. Information about the predictors (histopathology images and features derived thereof) was generally better reported, but still often missed key details which meant that it was unclear whether all tissue samples were processed similarly to avoid risks of bias from visual heterogeneity. \textcolor{black}{It was found that when patient characteristics were reported, they often showed a high risk of bias. Many studies included very small quantities of patients with specific differences from the majority (e.g. less than 20 patients with a different cancer subtype to the majority), causing a risk of spurious correlations and results which are not generalisable to the wider population.} 

\textcolor{black}{Reporting was particularly sparse in studies which used openly accessible data, possibly indicating that AI-focused researchers were not taking sufficient time to understand these datasets and ensure their research was clinically relevant. For example, many of the researchers who used TCGA data included frozen tissue sections without commenting on whether this was appropriate, despite the fact that pathologists do not consider them to be of optimal diagnostic quality. One paper handled TCGA data more appropriately, with a clear explanation of the positives and negatives of the dataset, and entirely separate models for FFPE and frozen slides \cite{Kalra2020}.}

\textcolor{black}{Sharing code can help to mitigate the effects of incomplete reporting and drastically improve reproducibility, but only 19 of the 45 papers did this, with some of these appearing to be incomplete or inaccessible. The better code repositories included detailed documentation to aid reproducibility, including environment set-up information \cite{Yu2020, Meng2023}, overviews of included functions \cite{Boehm2022, Lazard2022, Ho2023}, and code examples used to generate reported results \cite{Heindl2018}. }

\textcolor{black}{Two papers were found to have major discrepancies between the reported data and the study design, indicating much greater risks of bias than those seen in any other research \cite{Kasture2022,Ramasamy2023}. In one paper \cite{Kasture2022}, it was reported that TCGA-OV data was used for subtyping with 5 classes, despite this dataset only including high-grade serous and low-grade serous carcinomas. In the other paper \cite{Ramasamy2023}, it was reported that TCGA-OV data was used for slide-level classification into ovarian cancer and non-ovarian cancer classes using PAS-stained tissue, despite TCGA-OV only containing H\&E-stained ovarian cancer slides. }

\subsection*{Limitations of the Review}
\textcolor{black}{While the review protocol was designed to reduce biases and maximise the quantity of relevant research included, there were some limitations. This review is restricted to published literature in the English language, however, AI research may be published in other languages or made available as pre-prints without publication in peer-reviewed journals, making} this review incomplete. While most of the review process was completed by multiple independent researchers, the duplicate detection was performed by only a single researcher, raising the possibility of errors in this step of the review process, resulting in incorrect exclusions. Due to the significant time gap between the initial and final literature searches (approximately \textcolor{black}{12} months), there may have been inconsistencies in interpretations, both for data extraction and risk of bias assessments. Finally, this review focused only on light microscopy images of human histopathology samples relating to ovarian cancer, so may have overlooked useful literature outside of this domain. 


\subsection*{Development of the Field}
The field of AI in ovarian cancer histopathology diagnosis is rapidly growing, with more research published since the start of 2020 than in all preceding years combined. The earliest research, published between 2010-2013, used hand-crafted features to train classical ML methods such as SVMs. These models were used for segmentation \cite{Dong2010a,Dong2010b, Signolle2010,Janowczyk2012}, malignancy classification \cite{Janowczyk2011, Kothari2012}, grading \cite{Poruthoor2013}, and overall survival prediction \cite{Poruthoor2013}. Most of these early studies focused on IHC-stained tissue (5/7), which would be much less commonly used in subsequent research (\textcolor{black}{6/38}).

The field was relatively dormant in the following years, with only 6 papers published between 2014-2019, half of which had the same primary author \cite{BenTaieb2015,BenTaieb2016,BenTaieb2017}. These models still used traditional ML classifiers, though some used learned features rather than the traditional hand-crafted features. The models developed were used for histological subtyping \cite{BenTaieb2015,BenTaieb2016,BenTaieb2017} and cellular/tissue classification \cite{Lorsakul2017,Du2018,Heindl2018}. 

Since 2020 there has been a much greater volume of research published, most of which has involved the use of deep neural networks for automatic feature extraction and classification. 
Recent research has investigated a broader array of \textcolor{black}{diagnostic} outcomes, including the classification of primary cancer type \textcolor{black}{\cite{Kalra2020, Ramasamy2023}}, mutation status \textcolor{black}{\cite{Zeng2021, Nero2022, Ho2023}, homologous recombination deficiency status \cite{Lazard2022}, tumour-stroma reaction level \cite{Jiang2022}},  transcriptomic subtypes \cite{Yu2020, Zeng2021}, microsatellite instability \cite{Zeng2021}, and epithelial-mesenchymal transition status \cite{Hu2022}. \textcolor{black}{Three} additional \textcolor{black}{prognostic} outcomes have also been predicted in more recent literature - progression-free survival \textcolor{black}{\cite{Laury2021, Boehm2022, Yokomizo2022}, relapse  \cite{Nero2022, Yokomizo2022}, and treatment response \cite{Yaar2020, Yu2020, Wang2022a,Wang2022b, Wang2023}}. 

Despite progress within a few specific outcomes, there was no obvious overall trend in the sizes of datasets used over time, either in terms of the number of slides or the number of participants. Similarly, there was no evidence that recent research included more rigorous internal validations, 
though external validations have been increasing in frequency - no research before 2021 included any external validation with ovarian cancer data, but \textcolor{black}{seven studies published more recently did \cite{Shin2021, Zeng2021,Boschman2022, Farahani2022, Mayer2022, Wang2022b,Meng2023}.}  While these external validations were typically limited to small quantities of data, the inclusion of any external validation demonstrates progress from previous research. Such validations are essential to the clinical utility of these models as real-world implementation will require robustness to different sources of visual heterogeneity, with variation occurring across different data centres and within data centres over time. As this field continues to mature, we hope to see more studies conduct thorough validations with larger, high-quality independent datasets, including clearly reported protocols for patient recruitment and selection, pathology slide creation, and digitization. This will help to reduce the biases, limited reproducibility, and limited generalisability identified in most of the existing research in this domain.

\subsection*{Current Limitations and Future Recommendations}

A large proportion of published work did not provide sufficient clinical and pathological information to assess the risk of bias. 
It is important that AI researchers thoroughly report data provenance \textcolor{black}{to understand the extent of heterogeneity in the dataset, and to understand whether this has been appropriately accounted for in the study design. Modelling and analysis methods must also be thoroughly reported to improve reliability and reproducibility}. Researchers may find it useful to refer to reporting checklists, such as \emph{transparent reporting of a multivariable prediction model for individual prognosis or diagnosis} (TRIPOD) \cite{Collins2015}, to ensure that they have understood and reported all relevant details of their studies. \textcolor{black}{In many studies, it is not clear how AI would fit in the clinical workflow, or whether there are limitations in how these methods could be applied. AI researchers should ensure they understand the clinical context of their data and potential models before undertaking research to reduce bias and increase utility. Ideally, this will involve regular interactions with expert clinicians, including histopathologists and oncologists.}

\textcolor{black}{To further improve reproducibility, we recommend that researchers should make code and data available where possible. It is relatively easy to publish code and generate documentation to enhance usability, and there are few drawbacks to doing so when publishing research. Making data available is more often difficult due to data security requirements and the potential storage costs, but it can provide benefits beyond the primary research of the original authors. Digital pathology research in ovarian cancer is currently limited by the lack of openly accessible data, leading to over-dependence on TCGA, and causing many researchers to painstakingly collate similar but distinct datasets. These datasets often contain little of the heterogeneity seen in multi-centre, multi-scanner data, making it difficult for researchers to train robust models or assess generalisability. Where heterogeneous data is included, it often includes small quantities of data which are different to the majority, introducing risks of bias and confounding rather than helping to overcome these issues. TCGA-based studies are prone to this, with significant differences between TCGA slides originating from different data centres \cite{Dehkharghanian2023}, but with many of these centres only providing small quantities of data. Many researchers are reliant on open-access data, but there is a severe shortage of suitable open-access ovarian cancer histopathology data. Making such data available, with detailed protocols describing data creation, allows researchers to conduct more thorough analyses and significantly improve model generalisability and clinical implementability.}

\textcolor{black}{For AI to achieve clinical utility, it is essential that more robust validations are performed, especially considering the limitations of the available datasets. We} recommend that researchers should always conduct thorough analyses, using cross-validation, bootstrapping, and/or external validations to ensure that results are robust and truly reflect the ability of their model(s) to generalise to unseen data, and are not simply caused by chance. This should include reporting the variability of results (typically in a 95\% confidence interval), especially when comparing multiple models to help to distinguish whether one model is genuinely better than another or whether the difference is due to chance. Statistical tests can also be beneficial for these evaluations. Another option for capturing variability is Bayesian uncertainty quantification, which can be used to separate aleatoric (inherent) and epistemic (modelling) uncertainty. 

Current literature in this field can be largely characterised as model prototyping with homogeneous retrospective data. Researchers rarely consider the reality of human-machine interaction, perhaps believing that these models are a drop-in replacement for pathologists. However, these models perform narrow tasks within the pathology pipeline and \textcolor{black}{do not take into consideration the clinical context beyond their limited training datasets and siloed tasks}. We believe these models would be more beneficial (and more realistic to implement) as assistive tools for pathologists, providing secondary opinions or novel ancillary information. 
While current research is typically focused on assessing model accuracy without any pathologist input, different study designs could be employed to better assess the real-world utility of these models as assistive tools. For example, usability studies could investigate which models are most accessible and most informative to pathologists in practice, and prospective studies could quantify any benefits to diagnostic efficiency and patient outcomes, and investigate the robustness of models in practice. Understanding the effects of AI on the efficiency of diagnosis is particularly important given the limited supply of pathologists worldwide. As such, this type of research will significantly benefit clinical translation.



\newpage
\subsection*{Summary of recommendations}
\begin{itemize}
\setlength\itemsep{0.1em}
    \item Understand data and ensure planned research is clinically relevant before modelling, ideally involving clinicians throughout the project.
    \item Consider different study designs, including usability studies and/or prospective studies.
    \item Clearly report the context of any histopathology data, including how patients were recruited/selected, and how tissue specimens were processed to generate digital pathology images.
    \item Conduct thorough analyses using cross-validation, external validation, and/or bootstrapping.
    \item Make all code openly accessible (and data if possible).
\end{itemize}

\section*{Acknowledgments}
There was no direct funding for this research. JB is supported by the UKRI Engineering and Physical Sciences Research Council (EPSRC) [EP/S024336/1]. KA, PA are supported by the Tony Bramall Charitable Trust. AS is supported by Innovate UK via the National Consortium of Intelligent Medical Imaging (NCIMI) [104688], Cancer Research UK [C19942/A28832] and Leeds Hospitals Charity [9R01/1403]. The funders had no role in influencing the content of this research. For the purpose of open access, the author has applied a Creative Commons Attribution (CC BY) licence to any Author Accepted Manuscript version arising from this submission. 

\section*{Author Contributions}
JB created the study protocol with feedback and contributions from all other authors. JB, KA, KZ, NMO, and NR performed the risk of bias assessments. JB and KA performed data extraction. JB analysed extracted data and wrote the manuscript, with feedback and contributions from all other authors.

\section*{Competing Interests}
GH receives research funding from IQVIA. NMO receives research funding from 4D Path. All other authors declare no conflicts of interest.

\bibliography{main}

\newpage
\appendix

\section{Search strategy}\label{app:search}
Searches for all databases are shown here, with any text which is not directly input to the search bar in \textbf{bold} font. These searches are each a combination of three aspects - artificial intelligence, ovarian cancer, and histopathology. No filters were applied, and all options were left on their default settings. \textcolor{black}{The wildcard character, *, was used to search for multiple versions of the same word, for example, ``patholog*" searches for all of ``pathology", ``pathologist", ``pathologists", and ``pathological".}

\subsection{PubMed}
(“Machine Learning"[Mesh] OR “Artificial Intelligence"[Mesh] OR “Neural Networks, Computer"[Mesh] OR “support vector machine"[MeSH] OR “Deep Learning”[Mesh] OR “diagnosis, computer-assisted"[Mesh] OR “Machine learn$\ast$” OR “Artificial Intelligen$\ast$” 
OR (ML[Title/Abstract] NOT ($\mu$gml[Title/Abstract] OR $\mu$/ml[Title/Abstract] OR mgml[Title/Abstract] OR pgml[Title/Abstract] OR ngml[Title/Abstract] OR uiml[Title/Abstract] OR iuml[Title/Abstract] OR miuml[Title/Abstract] OR muiml[Title/Abstract] OR uml[Title/Abstract] OR gml[Title/Abstract] OR mlkg[Title/Abstract] OR milliliter$\ast$[Title/Abstract])) 
OR AI[Title/Abstract] OR “Computer Vision” OR “Neural network$\ast$” OR “Deep Network$\ast$” OR “Computer-aided Diagnosis” OR “Computer aided Diagnosis” OR Perceptron$\ast$ OR “Convolutional Network$\ast$” OR “Recurrent Network$\ast$” OR “Graph Network$\ast$” OR “Deep Learn$\ast$” OR “Deep-Learn$\ast$” OR Backprop$\ast$ OR “support vector$\ast$” OR ensemble$\ast$ OR “random forest$\ast$” OR “nearest neighbor$\ast$” OR “nearest neighbour$\ast$” OR “k-nearest neighbor$\ast$” OR “k-nearest neighbour$\ast$”  OR “Gradient boost$\ast$” OR “XGBoost$\ast$” OR “segmentation” OR “instance learning” OR “multi-instance learning” OR “Active Learning”)

AND (((ovar$\ast$ \textcolor{black}{OR fallopian}) AND (cancer$\ast$ OR mass$\ast$ OR carcinoma$\ast$ OR tumour$\ast$ OR tumor$\ast$ OR neoplasm$\ast$ OR malignan$\ast$ OR “carcinoma"[Mesh] OR “neoplasms"[Mesh])) OR “Ovarian Neoplasms”[Mesh] \textcolor{black}{OR “peritoneal cancer” OR “peritoneal carcinoma” OR “peritoneal tumo$\ast$”}) 

AND ((digit$\ast$ AND patholog$\ast$) \textcolor{black}{OR “computational patholog$\ast$”}  OR “tissue microarray$\ast$” OR histopath$\ast$ OR histolog$\ast$ OR “Whole Slide Imag$\ast$” OR “Tissue slide$\ast$” OR “pathology slide$\ast$” OR “pathology image$\ast$” OR Immunohistochem$\ast$ OR ((Haematoxylin OR Hematoxylin) AND Eosin) OR Histology[Mesh])

\subsection{Scopus}
TITLE-ABS-KEY(“Machine learn$\ast$” OR “Artificial Intelligen$\ast$” 
OR (“ML” AND NOT “$\ast$ $\mu$ ml” AND NOT “$\ast$g ml” AND NOT “$\ast$ui ml” AND NOT “$\ast$Ul ml” AND NOT “$\ast$iu ml” AND NOT “$\ast$u ml” AND NOT “$\ast$g ml” AND NOT “$\ast$ml kg” AND NOT milliliter$\ast$) 
OR AI OR “Computer Vision” OR “Neural network$\ast$” OR “Deep Network$\ast$” OR “Computer-aided Diagnosis” OR “Computer aided Diagnosis” OR Perceptron$\ast$ OR “Convolutional Network$\ast$” OR “Recurrent Network$\ast$” OR “Graph Network$\ast$” OR “Deep Learn$\ast$” OR “Deep-Learn$\ast$” OR Backprop$\ast$ OR “support vector$\ast$” OR ensemble$\ast$ OR “random forest$\ast$” OR “nearest neighbor$\ast$” OR “nearest neighbour$\ast$” OR “k-nearest neighbor$\ast$” OR “k-nearest neighbour$\ast$”  OR “Gradient boost$\ast$” OR “XGBoost$\ast$” OR “segmentation” OR “instance learning” OR “multi-instance learning” OR “Active Learning”)

AND TITLE-ABS-KEY(((ovar$\ast$ \textcolor{black}{OR fallopian}) AND (cancer$\ast$ OR mass$\ast$ OR carcinoma$\ast$ OR tumour$\ast$ OR tumor$\ast$ OR neoplasm$\ast$ OR malignan$\ast$)) \textcolor{black}{OR “peritoneal cancer” OR “peritoneal carcinoma” OR “peritoneal tumo$\ast$”})

AND TITLE-ABS-KEY((digit$\ast$ AND patholog$\ast$) \textcolor{black}{OR “computational patholog$\ast$”} OR “tissue microarray$\ast$” OR histopath$\ast$ OR histolog$\ast$ OR “Whole Slide Imag$\ast$” OR “Tissue slide$\ast$” OR “pathology slide$\ast$” OR “pathology image$\ast$” OR Immunohistochem$\ast$ OR ((Haematoxylin OR Hematoxylin) AND Eosin))

\subsection{Web of Science} 
(ALL=(“Machine learn$\ast$” OR “Artificial Intelligen$\ast$” OR “Computer Vision” OR “Neural network$\ast$” OR “Deep Network$\ast$” OR “Computer-aided Diagnosis” OR “Computer aided Diagnosis” OR Perceptron$\ast$ OR “Convolutional Network$\ast$” OR “Recurrent Network$\ast$” OR “Graph Network$\ast$” OR “Deep Learn$\ast$” OR “Deep-Learn$\ast$” OR Backprop$\ast$ OR “support vector$\ast$” OR ensemble$\ast$ OR “random forest$\ast$” OR “nearest neighbor$\ast$” OR “nearest neighbour$\ast$” OR “k-nearest neighbor$\ast$” OR “k-nearest neighbour$\ast$”  OR “Gradient boost$\ast$” OR “XGBoost$\ast$” OR “segmentation” OR “instance learning” OR “multi-instance learning” OR “Active Learning”) OR TS=(AI OR (“ML” NOT (“$\ast$ $\mu$ ml” OR “$\ast$g ml” OR “$\ast$ui ml” OR “$\ast$Ul ml” OR “$\ast$iu ml” OR “$\ast$u ml” OR “$\ast$g ml” OR “$\ast$ml kg” OR milliliter$\ast$))))

AND ALL=(((ovar$\ast$ \textcolor{black}{OR fallopian}) AND (cancer$\ast$ OR mass$\ast$ OR carcinoma$\ast$ OR tumour$\ast$ OR tumor$\ast$ OR neoplasm$\ast$ OR malignan$\ast$)) \textcolor{black}{OR “peritoneal cancer” OR “peritoneal carcinoma” OR “peritoneal tumo$\ast$”}) 

AND ALL=((digit$\ast$ AND patholog$\ast$) \textcolor{black}{OR “computational patholog$\ast$”}  OR “tissue microarray$\ast$” OR histopath$\ast$ OR histolog$\ast$ OR “Whole Slide Imag$\ast$” OR “Tissue slide$\ast$” OR “pathology slide$\ast$” OR “pathology image$\ast$” OR Immunohistochem$\ast$ OR ((Haematoxylin OR Hematoxylin) AND Eosin))

\subsection{Cochrane Central Register of Controlled Trials} 

\noindent \textbf{Search \#1}:

\textbf{All text}:
(“Machine learn$\ast$” OR “Artificial Intelligen$\ast$” OR “Computer Vision” OR “Neural network$\ast$” OR “Deep Network$\ast$” OR “Computer-aided Diagnosis” OR “Computer aided Diagnosis” OR Perceptron$\ast$ OR “Convolutional Network$\ast$” OR “Recurrent Network$\ast$” OR “Graph Network$\ast$” OR “Deep Learn$\ast$” OR “Deep-Learn$\ast$” OR Backprop$\ast$ OR “support vector$\ast$” OR ensemble$\ast$ OR “random forest$\ast$” OR “nearest neighbor$\ast$” OR “nearest neighbour$\ast$” OR “k-nearest neighbor$\ast$” OR “k-nearest neighbour$\ast$”  OR “Gradient boost$\ast$” OR “XGBoost$\ast$” OR “segmentation” OR “instance learning” OR “multi-instance learning” OR “Active Learning”) 

\vspace{0.5em}
\noindent \textbf{Search \#2}:

\textbf{Title-Abstract-Keyword}:
(“AI” OR (“ML” NOT (“$\ast$ $\mu$ ml” OR “$\ast$g ml” OR “$\ast$ui ml” OR “$\ast$Ul ml” OR “$\ast$iu ml” OR “$\ast$u ml” OR “$\ast$g ml” OR “$\ast$ml kg” OR milliliter$\ast$))) in Title Abstract Keyword

\vspace{0.5em}
\noindent \textbf{Search \#3}:

\textbf{All text}:
(((ovar$\ast$ \textcolor{black}{OR fallopian}) AND (cancer$\ast$ OR mass$\ast$ OR carcinoma$\ast$ OR tumour$\ast$ OR tumor$\ast$ OR neoplasm$\ast$ OR malignan$\ast$)) \textcolor{black}{OR “peritoneal cancer” OR “peritoneal carcinoma” OR “peritoneal tumo$\ast$”})

AND ((digit$\ast$ AND patholog$\ast$) \textcolor{black}{OR “computational patholog$\ast$”} OR “tissue microarray$\ast$” OR histopath$\ast$ OR histolog$\ast$ OR “Whole Slide Imag$\ast$” OR “Tissue slide$\ast$” OR “pathology slide$\ast$” OR “pathology image$\ast$” OR Immunohistochem$\ast$ OR ((Haematoxylin OR Hematoxylin) AND Eosin))

\vspace{0.5em}
\noindent \textbf{Final search}:

(\#1 OR \#2) AND \#3

\subsection{WHO-ICTRP}
((“Machine learn$\ast$” OR “Artificial Intelligen$\ast$” OR “Computer Vision” OR “Neural network$\ast$” OR “Deep Network$\ast$” OR “Computer-aided Diagnosis” OR “Computer aided Diagnosis” OR Perceptron$\ast$ OR “Convolutional Network$\ast$” OR “Recurrent Network$\ast$” OR “Graph Network$\ast$” OR “Deep Learn$\ast$” OR “Deep-Learn$\ast$” OR Backprop$\ast$ OR “support vector$\ast$” OR ensemble$\ast$ OR “random forest$\ast$” OR “nearest neighbor$\ast$” OR “nearest neighbour$\ast$” OR “k-nearest neighbor$\ast$” OR “k-nearest neighbour$\ast$”  OR “Gradient boost$\ast$” OR “XGBoost$\ast$” OR “segmentation” OR “instance learning” OR “multi-instance learning” OR “Active Learning”) 
OR (“AI” OR (“ML” NOT (“$\mu$/ml” OR “g/ml” OR “ui/ml” OR “Ul/ml” OR “iu/ml” OR “u/ml” OR “g/ml” OR “ml/kg” OR milliliter$\ast$))))

AND (((ovar$\ast$ \textcolor{black}{OR fallopian}) AND (cancer$\ast$ OR mass$\ast$ OR carcinoma$\ast$ OR tumour$\ast$ OR tumor$\ast$ OR neoplasm$\ast$ OR malignan$\ast$)) \textcolor{black}{OR “peritoneal cancer” OR “peritoneal carcinoma” OR “peritoneal tumo$\ast$”})

AND ((digit$\ast$ AND patholog$\ast$) \textcolor{black}{OR “computational patholog$\ast$”}  OR “tissue microarray$\ast$” OR histopath$\ast$ OR histolog$\ast$ OR “Whole Slide Imag$\ast$” OR “Tissue slide$\ast$” OR “pathology slide$\ast$” OR “pathology image$\ast$” OR Immunohistochem$\ast$ OR ((Haematoxylin OR Hematoxylin) AND Eosin))

\newpage
\section{Data extraction}\label{app:extract}

\begin{table}[htbp]
\centering
\rowcolors{2}{gray!15}{white}
 \begin{tabular}{|c|c|} 
 \hline
 \rowcolor{gray!30}
\textbf{Category} & \textbf{Fields}  \\
 \hline
 Overview & Internal ID. Lead author.   Year. Conference/Journal name. \\
 Data & \makecell{Number of development images. Total number of images. Type of samples. \\ FFPE/Frozen. Size of images. Tissue of origin. Number of development patients. \\ Total number of patients. Number of data collection centres. Type of stain. \\ Number of stainers. Scanners. Number of scanner types. Number of tissue \\ processing centes. Data origin countries. Number of pathologists for data labelling. \\ Online dataset. Prospective/retrospective. Clinical/research tissue. Data annotation. \\ Maximum magnification available. Supplementary datatypes. \\ Data exclusion reasons. Number of images excluded. Other cancer types included.} \\
 Methods & \makecell{Outcome. Outcome measure/classes. Outcome standards/definition. \\ Magnifications used. Patch sizes. Patches per image. Task type. Feature extraction \\ type. Feature extractors. AI in main method. Other AI methods. Optimiser. \\ Number of external validations. Differences to external validation set. Total external \\ validation images. Number of cross-validation folds. Number of non-novel methods \\ compared. Number of GPUs. Type of GPUs.} \\
 Results & \makecell{ \hspace{0.25cm}Internal test accuracy, error bounds. AUC, error bounds. Sensitivity/specificity, \hspace{0.25cm} \\ error bounds. Other metric 1, error bounds. Other metric 2, error bounds. \\ Other metric 3, error bounds. External training type. External test accuracy, \\ error bounds. AUC, error bounds. Sensitivity/specificity, error bounds. \\ Other metric 1, error bounds. Other metric 2, error bounds. Other metric 3, \\ error bounds. Type of error bounds. Model training time. Visualisation type.} \\
 Miscellaneous & Code availability. Data availability. Notes \\
 \hline
 \end{tabular}
 \caption{Summary of the fields used for data extraction. The full form is available at \protect\url{www.github.com/scjjb/OvCaReview}.}
 \label{table:extraction}
\end{table}

\end{document}